\documentclass{ws-procs975x65}

\def\beq{\begin{equation}}
\def\eeq{\end{equation}}

\begin{document}

\title{A search for chaos in the blazar: W2R 1926+42 and its possible consequence}
\author{Banibrata Mukhopadhyay}

\address{Department of Physics, Indian Institute of Science, Bangalore 560012\\
E-mail: bm@physics.iisc.ernet.in\\
}

\author{Rumen Bachev, Anton Strigachev} 

\address{Institute of Astronomy with National Astronomical Observatory, BAS, Bulgaria}



\begin{abstract}
We search for low-dimensional chaotic signatures in the optical lightcurve of the $Kepler$ field blazar W2R 1926+42. 
The frequently used correlation integral method is employed in our analysis. We find no apparent evidence for the presence 
of low-dimensional chaos in the lightcurve. If further confirmed, these results could be of importance for modeling the 
blazar emission mechanisms. 

\end{abstract}

\bodymatter


\section{Introduction}\label{intro}

Blazars are jet-dominated, highly variable active galactic nuclei (AGNs), powered by accretion and particle acceleration 
processes in the vicinity of a supermassive black hole. To be considered as a blazar, these objects should have their 
jets closely aligned with the line of sight. This perspective provides a unique opportunity to study in detail the 
physical processes within the jet, which is of significant importance not only for the astrophysics but also for the 
particle physics, as the energies reached within the jets exceed significantly those of the man-made accelerators.
However, the launching/emission mechanism of blazars is still not completely understood. 

In order to understand blazar, it is important to probe underlying accretion disk around the black hole, as
the jet is launched from the disk which is influenced by the black hole. However, the combined accretion-jet 
systems are nonlinear magnetohydrodynamical flows, which can be described, in general, by 10 
coupled differential equations:  3 momentum balance, 3 induction, 2 energy conservation, 1 mass conservation, 1 for no monopole.
Naturally, a question arises, whether the underlying nonlinear system is chaotic or stochastic. If it exhibits a 
low-dimensional chaos, then effectively a small number of independent variables are expected to control the system.
Hence, a model based on ``single engine" may be a perfect choice. Otherwise, the
models based on the idea of many independent zones are the likely scenario. 

The set of hydrodynamic equations describing one-temperature accretion-outflow systems was given by 
Ghosh and Mukhopadhyay \cite{gm09} and how it is to be modified in the presence of magnetic field 
in two-temperature flows can be understood from the works by Mukhopadhyay and Chatterjee \cite{mc15} and
Rajesh and Mukhopadhyay \cite{rm10}. As mentioned above, it forms altogether a set of 10 equations and,
hence, describing the evolution of 10 variables: 3 velocity components, 3 magnetic field components, 
ion and electron temperatures, pressure and density.
These 10 variables, in general, are expected to drive the entire observational appearance of the system, 
including its continuum variability. 
However, if the system is a low-dimensional (e.g. $\le 4$)
chaos, then all 10 variables effectively need not be independent to each other. Based on the 
$Keplar$ field data, here essentially we plan to explore the effective number of degrees of freedom 
determining the accretion-outflow system of blazars. This is will direct for correct modeling of blazars and, in general,
accretion-outflow systems.

In the present work, we apply the correlation/fractal dimension method to the optical lightcurve of a $Kepler$ field blazar 
W2R 1926+42, in order to find the number of the independent variables, driving the flow. The results might prove to 
be immensely helpful for the detailed modelling of the black hole accretion/jet flow.

\section{Computing correlation dimension}\label{model}

We follow Grassberger and Procaccia \cite{grass} and Karak, Dutta and Mukhopadhyay \cite{karak} to apply
correlation/fractal integral (CI) method, which is a statistical index 
measuring change in details of a system with the change in scale. The method is a suitable approach to reveal 
low-dimensional chaotic behavior in equally spaced experimental data sets (time series), such like blazar 
lightcurves \cite{Vio92}. The method was already applied for X-ray binaries: GRS~1915+105 (black hole),
Sco~X-1 (neutron star) etc. earlier. It is known that the transient source GRS~1915+105 exhibits 
12 different temporal classes \cite{belloni}. We explored the underlying nonlinear features, 
based on CI method,
in all 12 classes and found that some of them are chaotic and some other stochastic \cite{mukh,misra}. 
Furthermore, Sco~X-1 was shown to be chaotic \cite{karak} and finally by comparing nonlinear 
behaviors between black hole and neutron star sources, we argued heuristically that 
Cyg~X-3 is a black hole (whose nature is not quite confirmed yet).

The CI method is described in detail in the literature (see, e.g., Refs.~\refcite{Lehto,Bachev}) and 
relies on the construction of a new (empirical) phase space of embedding dimension $M$ from the available discrete 
data points. The data set (e.g. the lightcurve) is 
separated into segments of length $M$. Each segment can be considered as an $M$-dimensional vector 
($X_{i}$). Then the CI, expressed as
$$
C_M(r)=\dfrac{1}{N(N_c-1)} \sum_{i=1}^{N} \sum_{j=1,j\neq i}^{N} \Theta (r-|x_i-x_j|),
$$
is a measure of how the number of points in this $M$-dimensional space changes with the distance 
between any two them for each $M$. 
If the embedding dimension $M$ is larger than the effective number of degrees of freedom, i.e. 
the correlation dimension $D_2$ (which is basically the slop of $C_M(r)-r$ curves, as described below) 
of the dynamical system (the $attractor$), then saturation will 
occur, just like the number of random points on a flat surface will not grow faster than $r^{2}$, 
even though that from these 
points one can construct vectors of larger dimensions. The attractor/correlation dimension is then
$$
D_2(M) = \dfrac{d\log C_M(r)}{d\log r}.
$$ 

When applied to the famous Lorenz attractor 
$$
\dfrac{dx}{dt} = \sigma (y-x),~~\dfrac{dy}{dt} = x(\rho -z)-y,~~\dfrac{dz}{dt} = xy-\beta z,
$$ 
where see Ref.~\refcite{lorenz} for symbols and other details,
the CI method correctly computes the attractor dimension $D_2$ as about 2.04. The first integer number 
larger than the attractor dimension is the number of the independent variables (equations) driving the 
system (i.e. 3 in the case of Lorenz, as expected). 

\begin{figure}
\includegraphics[width=1.6in]{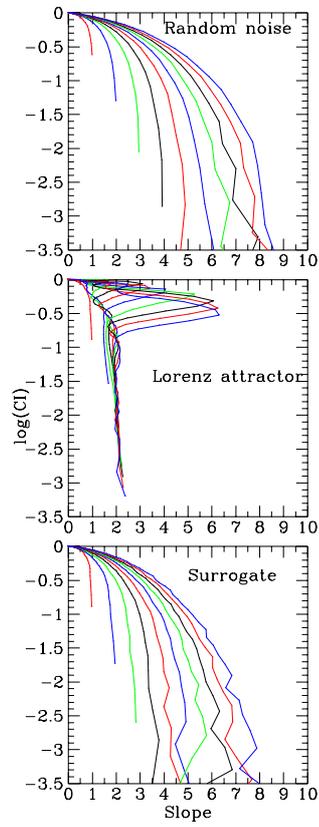}
\caption{CI diagrams for random noise (upper panel), Lorenz attractor (middle panel) and phase-randomized Lorenz attractor
surrogate (lower panel) \cite{Bachev}. Chosen different embedding dimensions,
from left to right curves, are  $1,2,3,...10$. 
}
\label{aba:fig1}
\end{figure}

A limitation to the CI method is the presence of periodic signals in the data set. Therefore it is always advisable to test as 
well surrogate data, produced from the original data set after phase randomization, thus preserving 
the original power spectrum and distribution. 
If the low-dimensional signature of CI is not destroyed after phase randomization, then it most probably is an artefact of the 
specific features of the power spectrum of the original data set. The presence of additional noise in the data can also conceal 
the low-dimensional attractor if such were present \cite{Lehto}.
Figure \ref{aba:fig1} shows that CIs 
of Lorenz data saturate with increasing $M$, while there is no saturation for its surrogate and 
random data, depicting how the ideal chaotic system should behave.

\section{Correlation dimension of the $Kepler$ lightcurve of the blazar W2R 1926+42}\label{LC}

\begin{figure}
\includegraphics[width=3in]{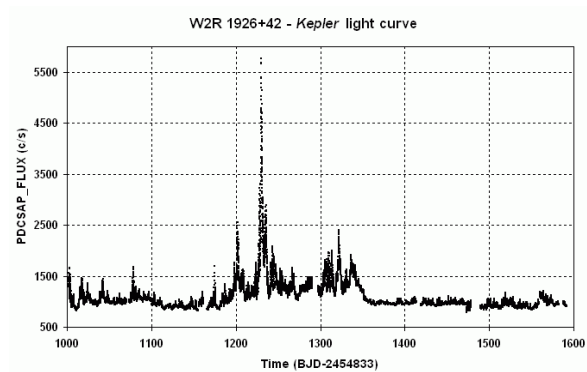}
\caption{The long term lightcurve of the blazar W2R 1926+42 from $Kepler$ satellite \cite{Bachev}.}
\label{aba:fig2}
\end{figure}

We choose the $Kepler$ field blazar due to the unprecedented high cadence, equal spacing and 
high photometric accuracy (rms of about 0.002) of the $Kepler$ satellite lightcurve data sets. The blazar was monitored 
in unfiltered optical light (which is where presumably the maximum of the synchrotron emission of the blazar is) 
for about 1.6 years, with time resolution of about 30 min. Figure \ref{aba:fig2} presents the entire lightcurve, publicly 
available from the $Kepler$ archive (details are in Bachev, Mukhopadhyay and Strigachev \cite{Bachev}).

\begin{figure}
\includegraphics[width=4in]{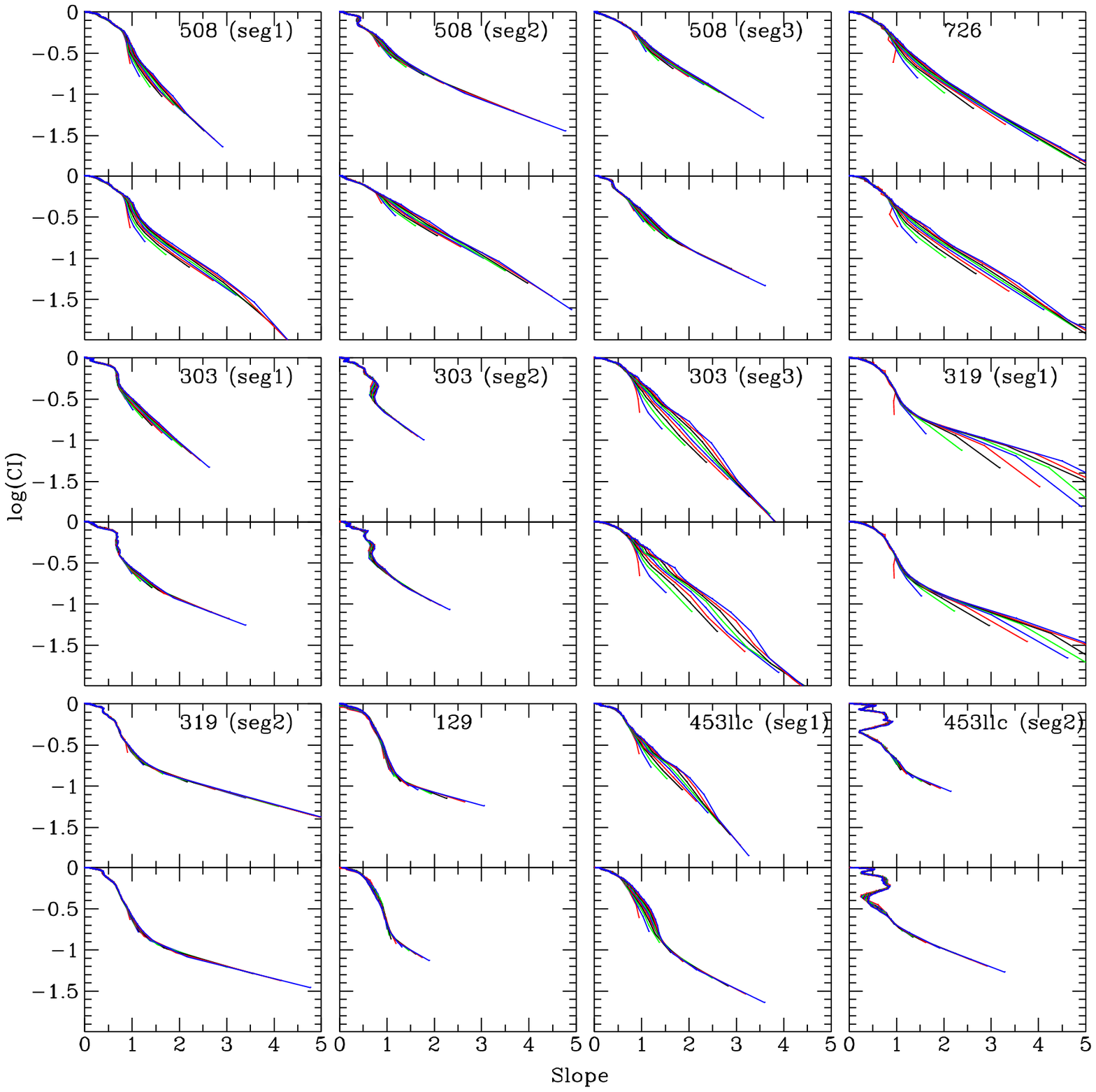}
\caption{CI diagrams, for different embedding dimensions same as those in Fig. \ref{aba:fig1}, 
for different segments of the blazar lightcurve. The last 3 digits of each data set 
(see $Kepler$ archive) are indicated. The
upper panel of each box is for the real data set, the lower one is for phase-randomized surrogate. None of the 
diagrams shows clear indication
for a strange attractor (chaos) of low dimension, especially taking into account that the surrogate diagrams 
are very similar to the ones of the real data \cite{Bachev}.}
\label{aba:fig3}
\end{figure}

Since the entire lightcurve of the blazar W2R 1926+42 consists of several segments with significant time gaps 
in between, we test each segment separately. The CI method has been 
applied to the surrogates of each segment as well. Figure \ref{aba:fig3} shows some of our results.


Although the pattern does not resemble the one of the pure random noise, we can not find any 
evidence for saturation (e.g. the presence of a low-dimensional attractor) in any of the segments 
for the first 10 embedding dimensions. Therefore, 
we find no signature of low-dimensional (e.g. $3-5$) attractor, which could have been useful 
to constrain/develop the theory. In addition, the surrogates show very similar patterns. 
The results, however, are somewhat similar to those reported by Lehto et al. \cite{Lehto} 
for their shot-noise model, even though they studied different type of objects and energy bands.

\section{Possible implications for the blazar jet models}\label{implications}

If further confirmed, our results (i.e. there is no low-dimensional behavior in the blazar lightcurves 
W2R 1926+42) indicate that the often invoked one-zone model may not correctly describe the blazar emission. 
The idea that synchrotron emission from a single 
emitting region and its temporal evolution should be controlled by a relatively small number of parameters, 
as the electron density distribution, magnetic field, Doppler factor, etc., cannot be confirmed by our analysis. 
Perhaps there are 
several, independently emitting/evolving regions that contribute to the blazar emission and thus the number of the 
governing variables can not be small.

\section{Summary}\label{summary}

Our analysis of a very long, well sampled lightcurve of the blazar W2R 1926+42 from the field of the $Kepler$ 
satellite reveals no apparent signatures of low-dimensional chaotic behavior.  These results, if 
further confirmed for this and other similar objects, suggest that perhaps the single zone models, 
often invoked to account for the blazar emissions could be 
oversimplification. Assuming the presence of many independently emitting regions, each evolving separately, 
is probably the 
correct approach to model the blazar spectral energy distribution and its temporal behavior.

\section*{Acknowledgments}
R.B. and A.S. would like to acknowledge the financial support from the National Science Fund grant DNTS 01/9, 2013 
(Indo-Bulgarian bilateral project). B.M. would thank an Indo-Bulgarian Project
funded by Department of Science and Technology, India, with grant no. INT/BULGARIA/P-8/12.


\begin{thebibliography}{0}

\bibitem{gm09} S. Ghosh and B. Mukhopadhyay, {\em RAA} {\bf 9}, 157 (2009).
\bibitem{mc15} B. Mukhopadhyay and K. Chatterjee, {\em Astrophys. J} {\bf 807}, 43 (2015).
\bibitem{rm10} S.R. Rajesh and B. Mukhopadhyay, {\em MNRAS} {\bf 402}, 961 (2010).
\bibitem{grass} P. Grassberger and I. Procaccia, {\em Phys. Rev. Lett.} {\bf 50}, 448 (1983).
\bibitem{karak} B. Karak, J. Dutta and B. Mukhopadhyay, {\em Astrophys. J} {\bf 708}, 862 (2010).
\bibitem{Vio92} R. Vio, S. Cristiani, O. Lessi and A. Provenzale, {\em Astrophys. J} {\bf 391}, 518 (1992).
\bibitem{belloni} T. Belloni, M. Klein-Wolt, M. M\'endez, M. van der Klis and
J. van Paradijs, {\em Astron. Astrophys.}  {\bf 355}, 271 (2000).
\bibitem{mukh} B. Mukhopadhyay, {\em AIP Conf. Proc.} {\bf 714}, 48 (2004).
\bibitem{misra} R. Misra, K.P. Harikrishnan, B. Mukhopadhyay, G. Ambika and A.K. Kembhavi, 
{\em Astrophys. J} {\bf 609}, 313 (2004).
\bibitem{Lehto} H. Lehto, B. Czerny and I. McHardy, {\em MNRAS} {\bf 261}, 125 (1993).
\bibitem{Bachev} R. Bachev, B. Mukhopadhyay and A. Strigachev, {\em Astron. Astrophys.} 
{\bf 576}, 17 (2015).
\bibitem{lorenz} E.N. Lorenz, {\em J. Atm. Sci.} {\bf 20}, 130 (1963).


\end{thebibliography}
\end{document}